\documentclass[useAMS]{mn2e}
\usepackage{graphicx}
\usepackage{amsfonts}

\title[]{Non-Gaussian velocity distributions  \\ -- The effect on virial mass estimates of galaxy groups}
\author[A.L.B. Ribeiro, P.A.A. Lopes and M. Trevisan]
{A.L.B. Ribeiro$^{1}$\thanks{E-mail: albr@uesc.br} P.A.A. Lopes$^{2}$ and M. Trevisan$^{3}$ \\
$^{1}$ Laborat\'orio de Astrof\'{\i}sica Te\'orica e Observacional, Universidade Estadual de Santa Cruz -- 45650-000, Ilh\'eus-BA, Brazil\\
$^{2}$ Observat\'orio do Valongo, Universidade Federal do Rio de Janeiro, Brazil\\
$^{3}$ Instituto Astron\^omico e Geof\'{\i}sico- USP, S\~ao Paulo-SP, Brazil}
\begin{document}

\date{Accepted 2011 February 22. Received 2011 February 6; in original form 2010 December 31.}

\pagerange{\pageref{firstpage}--\pageref{lastpage}} \pubyear{2010}

\maketitle

\label{firstpage}

\begin{abstract}
We present a study of 9 galaxy groups with evidence
for non-Gaussianity in their velocity distributions out to 4$R_{200}$. 
This sample is taken from 57 groups selected 
from the 2PIGG catalog of galaxy groups. Statistical analysis indicates
that non-Gaussian groups have masses significantly higher than Gaussian groups.
We also have found that all non-Gaussian systems
seem to be composed of multiple velocity modes. Besides, our results indicate that 
multimodal groups should be 
considered as a set of individual units with their own properties.
In particular, we have found that the mass distribution of such units
are similar to that of Gaussian groups. Our results reinforce the idea of non-Gaussian systems 
as complex structures in the phase space, likely corresponding to secondary infall
aggregations at a stage before virialization. The understanding of these objects is relevant for cosmological studies using groups and clusters through the mass function evolution.

\end{abstract}


\section{Introduction}

Groups of galaxies contain most of galaxies in the Universe and are the link
between individual galaxies and large-scale structures
(e.g., Huchra \& Geller 1982; Geller \& Huchra 1983; Nolthenius \& White 1987; Ramella et al. 1989). 
The dissipationless evolution of these systems is dominated by gravity.
Interactions over a relaxation time tend to distribute the velocities of the galaxy members
in a Gaussian distribution (e.g. Bird \& Beers 1993).
Thus, a way to access the dynamical stage of galaxy groups is to study their velocity distributions.
Evolved systems are supposed to have Gaussian velocity distributions, while those with
deviations from normality are understood as less evolved systems.
Hou et al. (2009) have examined three goodness-of-fit tests (Anderson-Darling, Kolmogorov and $\chi^2$ tests) to find which statistical tool is best able to distinguish between relaxed and non-relaxed galaxy groups. Using Monte Carlo simulations and
a sample of groups selected from the CNOC2, they found that the Anderson-Darling (AD) test
is far more reliable at detecting real departures from normality in small samples. Their results show
that Gaussian and non-Gaussian groups present distinct velocity dispersion profiles, suggesting that discrimination
of groups according to their velocity distributions may be a promising way to access 
the dynamics of galaxy systems. 

Recently, Ribeiro, Lopes \& Trevisan (2010) extended up this kind of analysis to the outermost edge of groups
to probe regions where galaxy systems  might not be in dynamical equilibrium. They found significant segregation effects after 
splitting up the sample in Gaussian and non-Gaussian systems. In the present work, we try to further understand the nature of
non-Gaussian groups. In particular, we investigate the
problem of mass estimation for this class of objects, and their behaviour in the phase space.
The paper is organized as follows: in Section 2 we present data and methodology;
Section 3 contains a statistical analysis with sample tests and multimodality diagnostics for 
non-Gaussian groups; finally, in Section 4 we summarize
and discuss our findings.

\section{Data and Methodology}

\subsection{2PIGG sample}

We use a subset of the 2PIGG catalog,
corresponding to groups located in areas
of at least 80\% redshift coverage in 2dF data out to 10 times the
radius of the systems, roughly estimated from the projected harmonic mean (Eke et al. 2004).
The idea of working with such large areas is to probe the effect of
secondary infall onto groups. Members and interlopers were redefined after the
identification of gaps in the redshift distribution according
to the technique described by Lopes et al. (2009). Finally, a virial 
analysis is perfomed to estimate the groups' properties.
See details in Lopes et al. (2009a); Ribeiro et al. (2009); and Ribeiro, Lopes \& Trevisan (2010).
We have classified the groups after applying the AD test  to their galaxy velocity
distributions (see Hou et al. 2009 for a good description of the test). 
This is done for different distances, producing the following
ratios of non-Gaussian groups: 6\% ($R \leq 1R_{200}$), 9\% ($R\leq 2R_{200}$), 
and 16\% ($R\leq 3R_{200}$ and $R\leq 4R_{200}$). We assume this latter
ratio (equivalent to 9 systems) as the correct if one desires to extending up the analysis to the regions where galaxy groups might not be in dynamical equilibrium.
Approximately 90\% of all galaxies
in the sample  have distances $\leq 4R_{200}$. This is the natural cutoff in space we have made in this work. 
Some properties of galaxy groups are presented in Table 1,
where non-Gaussian groups are identified with an asterisk.
Cosmology is defined by $\Omega_m$ = 0.3, $\Omega_\lambda$ = 0.7, and $H_0 = 100~h~{\rm km~s^{-1}Mpc^{-1}}$
Distance-dependent quantities are calculated using $h=0.7$.

\begin{table}
\caption{Main properties of groups}           
\label{tab1}      
\small{              
\begin{tabular}{l c c c c c}       
\hline\hline   
Group          & $R_{200}$ (Mpc) & $M_{200}$ ($10^{14}~M_{\sun}$)& $\sigma $ (km/s) & $N_{memb}$ & $N_{200}$ \\ 
\hline 
55&0.689&0.400&173.203&16&9\\
60&0.664&0.359&120.395&41&12\\
84&1.025&1.326&224.715&54&10\\
91&0.848&0.754&164.293&34&4\\
102$^\ast$&1.550&4.602&433.218&32&8\\
130&0.999&1.245&240.908&43&11\\
138$^\ast$&1.271&2.578&348.885&74&14\\
139&0.997&1.250&235.740&39&12\\
169&0.765&0.573&225.212&9&4\\
177$^\ast$&1.178&2.100&336.309&45&14\\
179$^\ast$&1.663&5.897&471.352&23&10\\
181&1.247&2.488&393.379&29&12\\
188&0.514&0.174&153.714&8&2\\
191&0.707&0.456&268.517&14&10\\
197&0.897&0.930&297.903&13&7\\
204&1.493&4.333&473.813&33&12\\
209&0.680&0.407&132.095&15&4\\
222&1.243&2.494&315.355&22&8\\
236&1.258&2.575&331.308&19&7\\
271&0.968&1.093&240.691&41&15\\
326&0.648&0.333&167.764&21&7\\
352$^\ast$&1.656&5.618&496.694&31&13\\
353&0.873&0.826&220.025&34&12\\
374&0.902&0.919&217.038&21&10\\
377&0.480&0.138&132.466&13&5\\
387&0.649&0.345&125.308&32&5\\
398&0.781&0.601&182.800&29&12\\
399&0.917&0.973&183.397&34&7\\
409&1.078&1.583&251.071&64&11\\
410&0.847&0.769&199.465&42&8\\
428&1.201&2.191&342.440&26&14\\
435&0.606&0.283&123.549&23&4\\
444&0.820&0.697&244.491&13&7\\
447$^\ast$&1.169&2.032&296.919&37&6\\
453&0.827&0.720&271.840&17&10\\
455&1.588&5.088&454.227&65&15\\
456&0.319&0.041&58.138&16&3\\
458&0.642&0.336&115.674&18&4\\
466&1.507&4.381&554.511&23&13\\
471&1.128&1.832&399.078&22&14\\
475$^\ast$&1.152&1.952&294.413&28&6\\
479&0.806&0.671&160.887&17&6\\
480&0.923&1.006&272.058&23&12\\
482&0.663&0.373&202.903&17&9\\
484&1.099&1.702&268.116&32&7\\
485&1.679&6.076&559.140&29&18\\
488&1.220&2.334&326.566&42&12\\
489&0.872&0.852&188.283&30&6\\
493&1.170&2.061&323.268&30&10\\
504&1.319&2.985&429.207&30&13\\
505$^\ast$&1.176&2.111&336.804&20&7\\
507$^\ast$&1.459&4.036&436.498&27&11\\
513&0.724&0.495&178.637&19&6\\
515&1.117&1.817&298.220&30&13\\
519&0.808&0.689&207.011&17&4\\
525&0.495&0.158&108.230&9&3\\
536&2.003&10.576&676.902&45&24\\
\hline
\end{tabular}
}
\end{table}

\section{Non-Gaussianity and Mass}

\subsection{The mass bias}

We consider the two sample statistical problem for the Gaussian (G) and NG (non-Gaussian) subsamples.
We choose  the mass resulting from the virial analysis as the property to illustrate
the comparison between the subsamples. 
Both Kolmogorov-Smirnov (KS) and Cramer-von Mises (CvM) tests reject the
hypothesis that the NG subsample is distributed as the G subsample, with p-values 0.00001 and 0.00029, 
respectively. For these tests, we have used 1000 bootstrap replicas of each subsample to alleviate the
small sample effect. The result indicates an inconsistency between the mass distributions of
G and NG groups. This could represent a real physical difference or, more probably, an indication of
a significant bias to higher masses in NG groups. The median mass
for this subsample is $\langle M_{200}^{NG}\rangle=2.57\times 10^{14}~M_{\sun}$, while it is
$\langle M_{200}^{G}\rangle=8.85\times 10^{13}~M_{\sun}$ for the G subsample, thus 
$\langle M_{200}^{NG}\rangle$ is larger by a factor
of $\sim 2.9$. In the following, we investigate this mass bias looking for
features in the velocity distributions of galaxy groups.

\subsection{Exploring non-Gaussianity}

The shape of velocity distributions may reveal a signature of the
dynamical stage of galaxy groups. For instance, systems with heavier tails than
predicted by a normal parent distribution may be contamined by interlopers. Otherwise, systems with
lighter tails than a normal may be multimodal, consisting of overlapping distinct populations
(e.g. Bird \& Beers 1993). 
We now try to understand the mass bias in the
NG subsample by studying non-Gaussianity in the velocity distributions.
Since we have carefully removed interlopers from each field (see Section 2.1; and Lopes et al. 2009a), 
we consider here that the most probable cause of normality deviations
in our sample is due to a superposition of modes in
the phase space (e.g. Diemand \& Kuhlen 2008). Visual inspection of
radial velocity histograms of NG systems suggests that
multipeaks really happen in most cases (see Figure 1).

We statistically check multimodality
by assuming the velocity distributions as Gaussian mixtures with unknown
number of components. We use the Dirichlet Process mixture (DPM) model to
study the velocity distributions. The DPM model is a Bayesian
nonparametric methodology that relies on Markov Chain Monte Carlo
(MCMC) simulations for exploring mixture models with an unknown number of components (Diebolt \& Robert 1994). 
It was first formalised in Ferguson (1973) for general Bayesian statistical
modeling. The DPM is  a distribution over  $k$-dimensional 
discrete distributions, so each draw from a Dirichlet process is itself a distribution.
Here, we assume that a galaxy group is a set of $k$ components, $\sum_{i=1}^k \pi_i f(y|\theta_i)$,
with galaxy velocities distributed according to Gaussian distributions with mean and variance unknown.
In this framework, the numbers $\pi_i$ are the mixture
coefficients that are drawn from a Dirichlet distribution.
In the DPM model, the actual number of components $k$ used to model
data is not fixed, and can be automatically inferred from data using the usual
Bayesian posterior inference framework. See Neal (2000) for a survey of MCMC inference
procedures for DPM models. 

In this work, we find $k$ using
the R language and environment (R Develoment Core Team) under the dpmixsim library
(da Silva 2009). The code  implements mixture models with normal structure (conjugate normal-normal DPM model). 
First, it finds the coefficients $\pi_i$, and then separates the components of the
mixture, according to the most probable values of $\pi_i$,
in the distributional space, leading to a partition of this space into regions (da Silva 2009).
The results can be visually analysed by plotting the estimated kernel densities for the
MCMC chains.  In Figure 2, we show the DPM diagnostics for each group, that is, the deblended
modes in the velocity distributions. We have found the following number of modes
per group: 4/102, 3/138, 3/177, 2/179, 5/352, 3/447, 2/475, 3/505, 3/507.
Therefore, all non-Gaussian groups in our sample are multimodal (reaching a total of 28 modes) 
according to the DPM analysis.
Unfortunately, we cannot compute
the physical properties of 13 modes, due to intrinsic scattering in velocity data (and/or to the smallness
of the modes -- those with less than 4 members). The properties of the other 15 modes are presented in Table 2.

\subsection{The mass bias revisited}

\begin{figure}
\includegraphics[width=84mm]{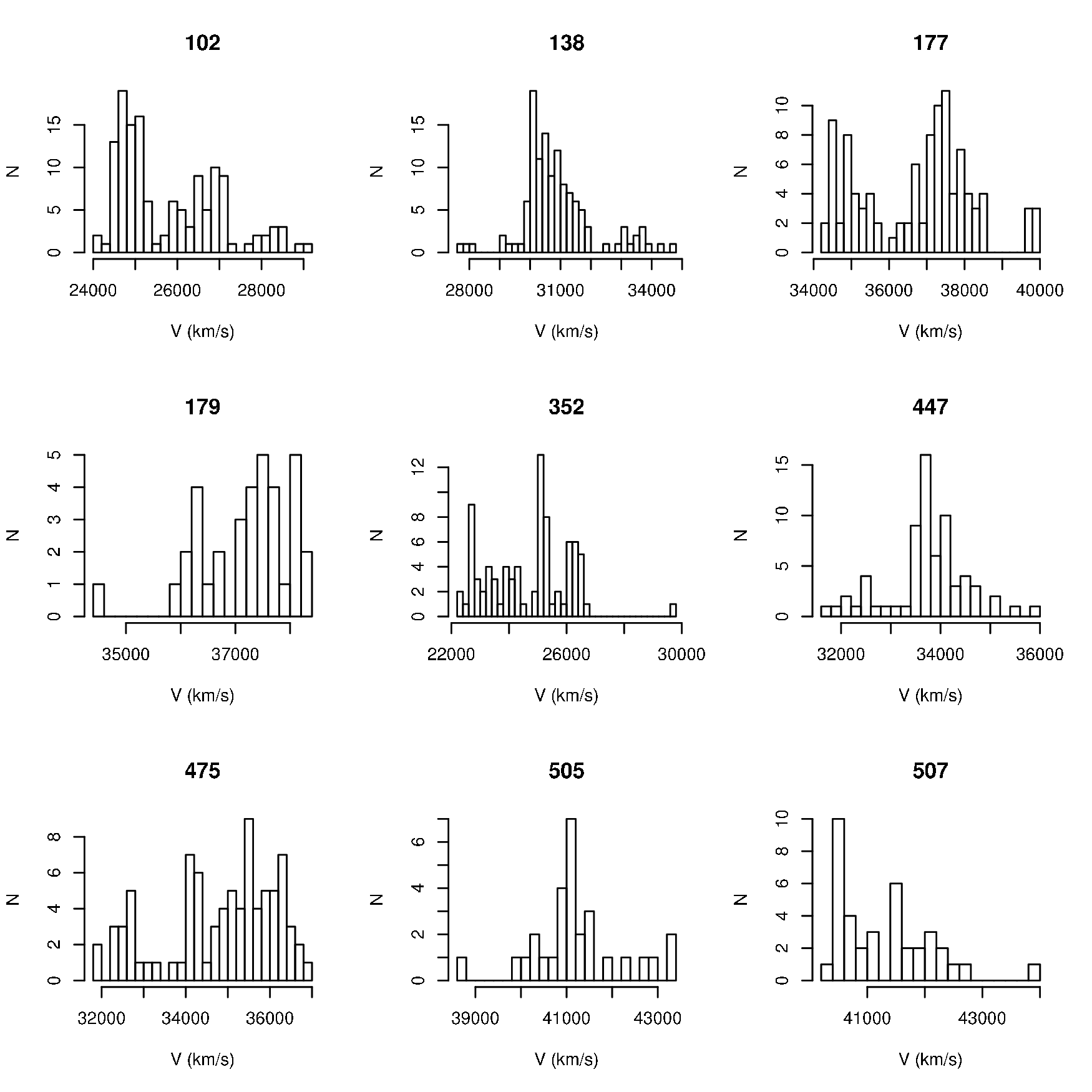}
\caption{Histograms of radial velocities for all non-Gaussian groups.}
\label{}
\end{figure}

Now, we perform again the statistical tests, comparing the distribution of mass 
in NG and G with the new sample (see Figure 3).
First, we compare the NG subsample and the sample of modes (M).
Both KS and CvM tests reject the hypothesis that M is distributed as NG,
with p=0.0211 and p=0.0189, respectively. Then, we compare G and M.
Now, the tests accept the hypothesis that M is distributed as G, with
p=0.4875 and p=0.4695. Hence, the distribution of the modes deblended from
non-Gaussian groups are themselves mass distributed as Gaussian groups. Also, the
median mass of the M sample is $\langle M_{200}^{M}\rangle=1.33\times 10^{14}~M_{\sun}$,
a value larger than $\langle M_{200}^{G}\rangle$ only by a factor $\sim 1.5$ (before
deblending groups the factor was $\sim$ 2.9).
This consistency between G and M objects indicate that non-Gaussian groups are
a set of smaller systems, probably forming an aggregate out of equilibrium.
Indeed, groups are the link between galaxies and larger structures. Thus,
our results suggest we are witnessing secondary infall (the secondary modes) onto
a previously formed (or still forming) galaxy system (the principal mode).
Naturally, we cannot discard the possibility of NG groups being unbound systems of smaller groups seen in projection,
although the properties of the independent modes are quite
similar to those found in physically bound groups.

\begin{figure}
\includegraphics[width=84mm]{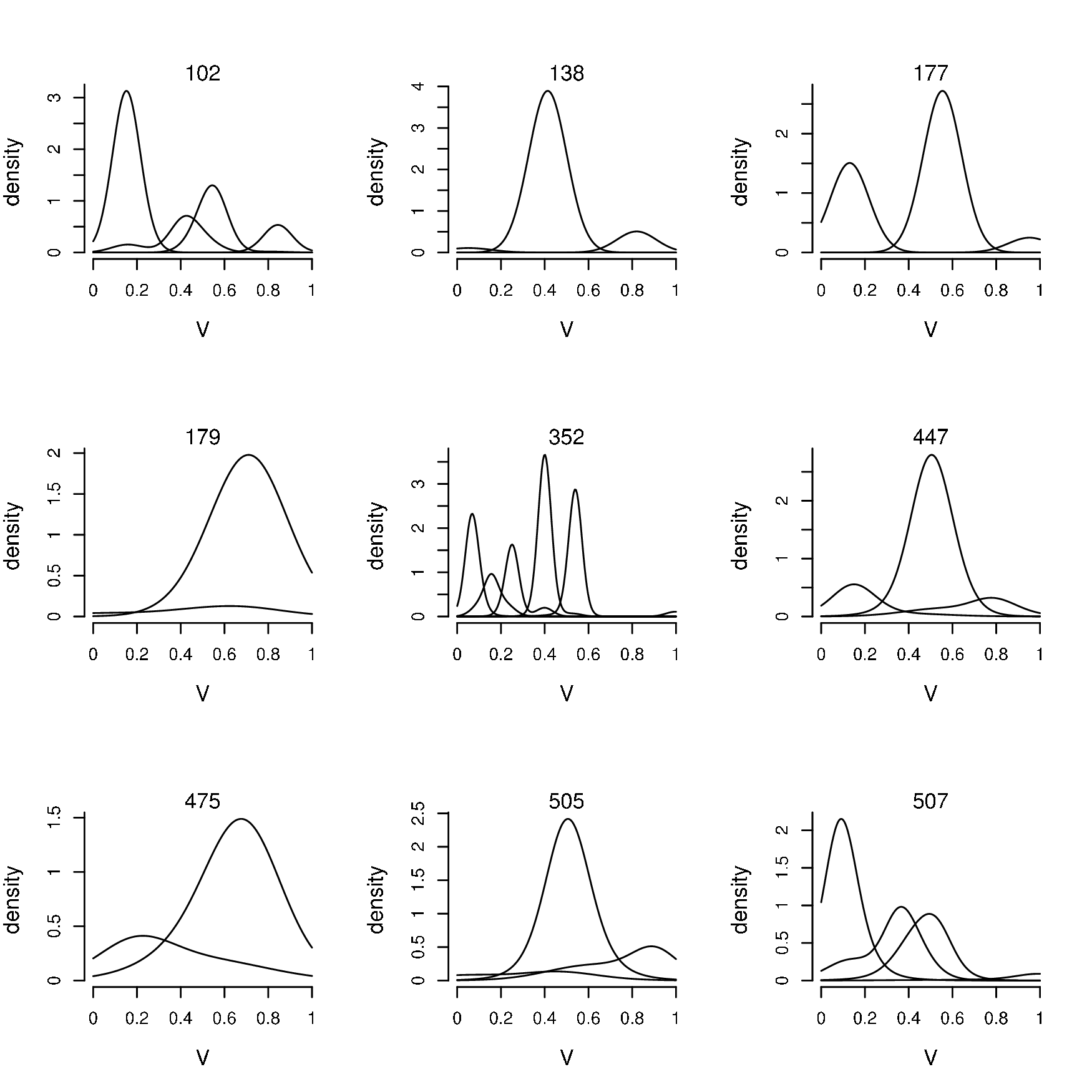}
\caption{DPM density probability decomposition. Velocities are rescaled to the [0,1] interval.}
\label{}
\end{figure}

\begin{table}
\caption{Main properties of individual modes}           
\label{tab1}      
\small{              
\begin{tabular}{l c c c c c}       
\hline\hline   
Mode          & $R_{200}$ (Mpc) & $M_{200}$ ($10^{14}~M_{\sun}$)& $\sigma $ (km/s) & $N_{200}$ \\ 
\hline 
102$_a$ & 1.077 & 1.549 & 337.482 & 21\\
102$_b$ & 0.728 & 0.477 & 153.372 & 4\\
138$_a$ & 1.550 & 4.677 & 498.927 & 35\\
177$_a$ & 1.249 & 2.481 & 369.569 & 11\\
177$_b$ & 1.212 & 2.289 & 362.462 & 14\\
179$_a$ & 1.326 & 3.001 & 336.265 & 8\\
352$_a$ & 0.487 & 0.142 & 158.571 & 6\\
352$_b$ & 0.941 & 1.033 & 218.210 & 6\\
352$_c$ & 0.697 & 0.416 & 208.670 & 3\\
352$_d$ & 0.558 & 0.213 & 181.206 & 5\\
447$_a$ & 1.016 & 1.330 & 240.437 & 3\\
447$_b$ & 1.192 & 2.155 & 441.382 & 5\\
475$_a$ & 1.500 & 4.308 & 470.516 & 19\\
475$_b$ & 2.345 & 16.347 & 829.977 & 10\\
505$_a$ & 0.928 & 1.040 & 228.411 & 3\\
\hline
\end{tabular}
}
\end{table}

\begin{figure}
\includegraphics[width=84mm]{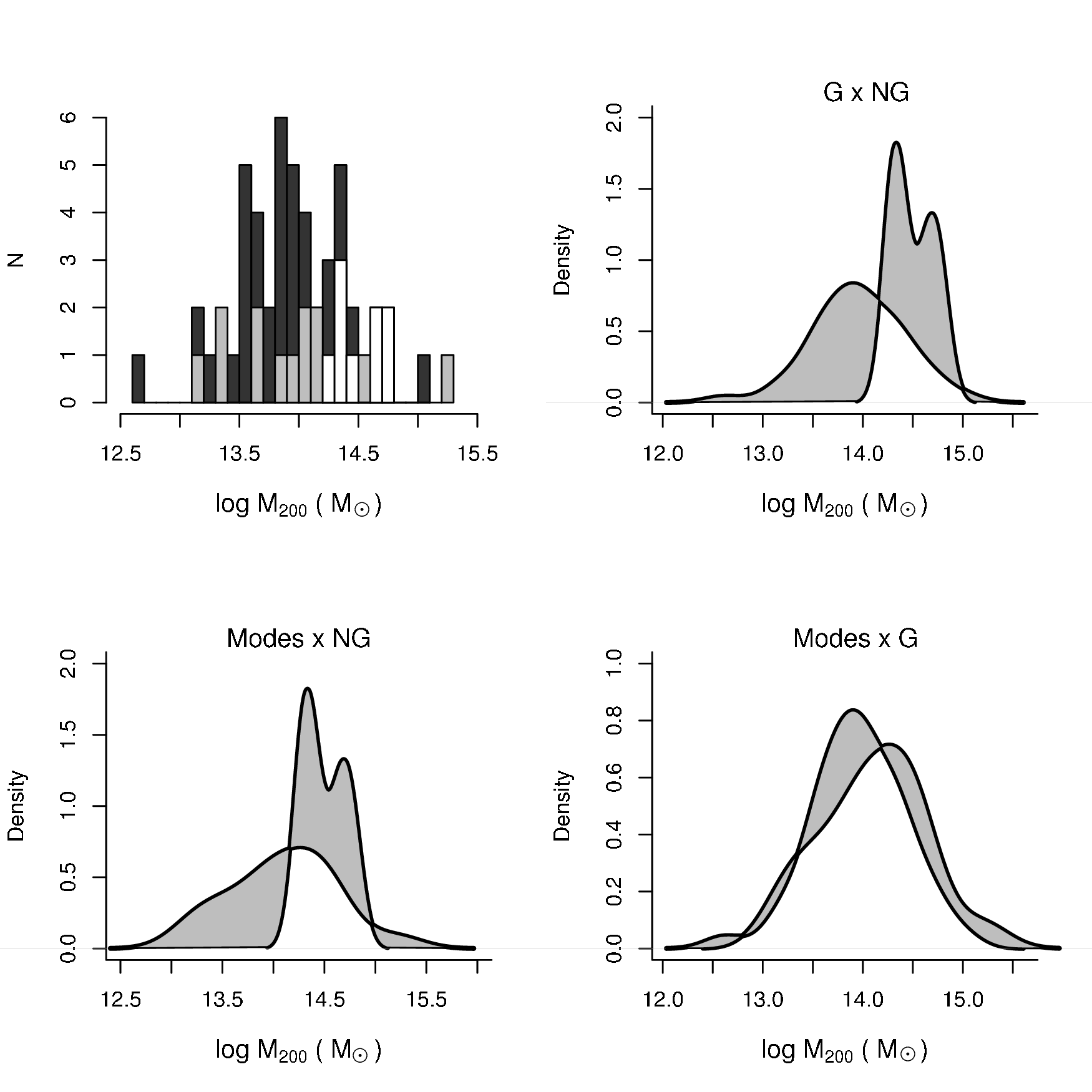}
\caption{Histograms for Gaussian (dark gray) and non-Gaussian (white) groups.
Histogram of modes is plotted in light gray. Density probability comparison among systems. Intersections areas are in white.}
\label{}
\end{figure}

\section{Discussion}

We have classified galaxy systems after applying the AD normality test to their 
velocity distributions up to the outermost edge of the groups. The purpose was to
to investigate regions where galaxy systems  might not be in dynamical equilibrium.
We have studied  57 galaxy groups selected from the 2PIGG catalog (Eke et al. 2004) 
using 2dF data out to $4R_{200}$. This means we probe galaxy distributions near to the
turnaround radius, thus probably taking into account all members in the infall pattern around the groups
(e.g. Rines \& Diaferio 2006; Cupani, Mezzetti \& Mardirossian 2008). 
The corresponding velocity fields depend on the local density of matter. High density regions should
drive the formation of virialized objects, whereas low density environments are more likely to
present streaming motions, i.e., galaxies falling toward larger potential wells
constantly increasing the amplitude of their clustering strength (e.g. Diaferio \& Geller 1997).

We have found that 84\% of the
sample is composed of systems with Gaussian velocity distributions. 
These systems could result from  the collapse and virialization of high density regions with
not significant secondary infall. They could be groups surrounded by well organized infalling motions, possibly
reaching virialization at larger radii. Theoretically, in regions outside $R_{200}$, we should apply the non-stationary
Jeans formalism, leading us to the virial theorem with some correction terms. These are due to the infall velocity gradient along
the radial coordinate and to the acceleration of the mass accretion process (see Cupani 2008). Their contribution is likely to be generally
negligible in the halo core where the matter is set to virial equilibrium and to become significant in the halo outskirts where the matter is still accreting (see e.g. Cupani, Mezzetti \& Mardirossian 2008). However, a slow and well organized infalling motion to the center of the groups could diminish the importance of the correction terms, i.e., the systems could be in a quasi-stationary state outside $R_{200}$. The high fraction of groups immersed in such surroundings suggests that ordered infalling motion around
galaxy groups might happen quite frequently in the Universe.

In any case, Gaussian groups can be considered as dynamically more evolved systems (see Ribeiro, Lopes \& Trevisan 2010). The remaining 16\% of the sample is composed of non-Gaussian groups.
We found that  these sytems have masses significantly larger than Gaussian groups.
This biasing effect in virial masses is basically due to the higher velocity dispersions in NG groups.
Ribeiro, Lopes \& Trevisan (2010) found that NG groups have rising velocity dispersion profiles. A similar result was found by Hou et al. (2009). Rising velocity dispersion profiles could be related to the higher fraction of blue galaxies in the outskirts of some galaxy systems (see Ribeiro, Lopes \& Trevisan 2010; and Popesso et al. 2007).

At the same time, the NG subsample is  
composed of multipeak objects, identified by the DPM model analysis applied to the velocity space.
These results indicate that secondary infall might be biasing the mass estimates of these  groups.
Thus, the NG systems could result from the collapse of less dense regions with
significant secondary infall input. Contrary to Gaussian groups, the surroundings of NG systems do not seem to
be in a quasi-stationary state. They are dynamically complex. Actually, we have found these groups can be modelled as assemblies
of smaller units. After deblending groups into a number of individual modes, we have verified
that the mass distribution of these objects is consistent with that of Gaussian groups,
suggesting that each unit is probably a galaxy group itself, a system formed during the streaming motion
toward the potential wells of the field.

\begin{figure}
\includegraphics[width=84mm]{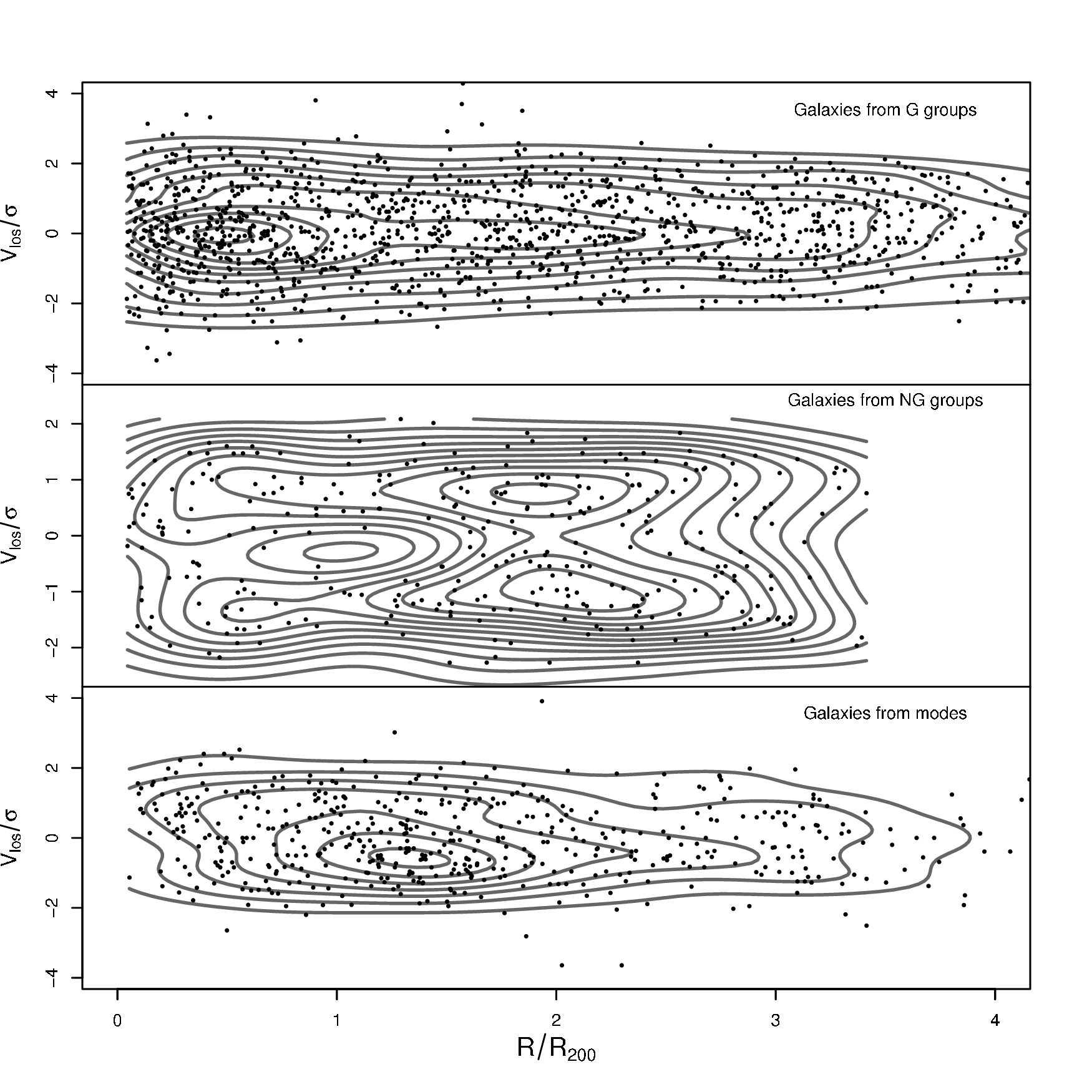}
\caption{Phase space diagrams for a typical Gaussian (upper box), non-Gaussian (middle box) and modes (lower box) systems. Density contours indicate galaxy concentration across the diagrams. Distances to the center of the systems are
normalized by $R_{200}$. Radial velocities are subtracted from the median velocity and divided by the
velocity dispersion of the groups.}
\label{}
\end{figure}

Our results reinforce the idea of NG systems as  complex structures in the phase space 
out to $4R_{200}$. This scenario is illustrated in Figure 4, where we show the stacked G and NG groups, and the stacked
modes. Galaxies in these composite groups have distances and line-of-sight velocities with respect to the
centers normalized by $R_{200}$ and $\sigma$, respectively. 
In the upper box, we present the G stacked group.
Note that galaxies are extensively concentrated in the phase space diagram, with a single density peak
near to  0.5$R_{200}$, revealing a well organized system around this point. 
Also, note that the density peak has
$V_{los}/\sigma\approx 0$. Contour density lines suggest that ordered shells of matter are moving toward the
center.  A different result is found for the NG stacked group.
In the middle box of Figure 4, we see a less concentrated galaxy distribution, with less tight density
contour levels, presenting a density peak slightly larger than $R_{200}$, and  two additional peaks, around
2$R_{200}$, possibly interacting with the central mode. The additional peaks are not aligned at $V_{los}/\sigma\approx 0$, suggesting a less symmetrical galaxy distribution in the phase space.
These features suggest that non-Gaussian systems are distinct, and dynamically younger than Gaussian groups,
which agrees well with the results of Ribeiro, Lopes \& Trevisan (2010). Also in Figure 4, we present
the modes stacked system in the lower box. Similar to the Gaussian case, we have a single peak in the phase space, near to 1.5$R_{200}$. Galaxy distribution however is still less symmetrical than in Gaussian groups, with the peak not aligned at $V_{los}/\sigma\approx 0$. This suggests that, although more organized than NG systems, modes are probabably dynamically distinct
and younger than Gaussian groups as well.

Our work points out the importance of
studying NG systems both to possibly correct their mass estimates and multiplicity functions, as well as to better understand galaxy clustering at group scale. Understanding of these objects is also relevant for cosmological studies using groups and clusters through the evolution of the mass function (Voit 2005). Using systems with overestimated properties may lead to a larger scatter in the mass calibration (Lopes et al. 2009b) and could also affect the mass function estimate (Voit 2005).

\section*{Acknowledgments}
We thank the referee for raising interesting points.
We also thank A.C. Schilling and S. Rembold for helpful discussions.
ALBR thanks the support of CNPq, grants 306870/2010-0 and 478753/2010-1. 
PAAL thanks the support of FAPERJ, process 110.237/2010. MT thanks the support of FAPESP, process 2008/50198-3.

\bsp

\label{lastpage}

\end{document}